\documentclass{aa}  
\usepackage{booktabs}

\usepackage{array}
\usepackage{natbib,twoopt}
\usepackage[breaklinks=true]{hyperref} 
\bibpunct{(}{)}{;}{a}{}{,}             
\makeatletter
  \newcommandtwoopt{\citeads}[3][][]{\href{http://adsabs.harvard.edu/abs/#3}%
    {\def\hyper@linkstart##1##2{}%
     \let\hyper@linkend\@empty\citealp[#1][#2]{#3}}}
  \newcommandtwoopt{\citepads}[3][][]{\href{http://adsabs.harvard.edu/abs/#3}%
    {\def\hyper@linkstart##1##2{}%
     \let\hyper@linkend\@empty\citep[#1][#2]{#3}}}
  \newcommandtwoopt{\citetads}[3][][]{\href{http://adsabs.harvard.edu/abs/#3}%
    {\def\hyper@linkstart##1##2{}%
     \let\hyper@linkend\@empty\citet[#1][#2]{#3}}}
  \newcommandtwoopt{\citeyearads}[3][][]%
    {\href{http://adsabs.harvard.edu/abs/#3}
    {\def\hyper@linkstart##1##2{}%
     \let\hyper@linkend\@empty\citeyear[#1][#2]{#3}}}
\makeatother
\usepackage{graphicx}
\usepackage{txfonts}
\usepackage{color}

\begin{document} 

   \title{The Rossiter-McLaughlin effect and exoplanet transits: A delicate association at medium and low spectral resolution.} 
   \titlerunning{Low/Medium-resolution Rossiter-McLaughlin effect}

   \author{Yann Carteret
          \inst{1},
          Vincent Bourrier\inst{1}
          \and
          William Dethier\inst{2}
          } 

   \institute{Observatoire Astronomique de l’Université de Genève, Chemin Pegasi 51b, CH-1290 Versoix, Switzerland\\
              \email{yann.carteret@gmail.com}
         \and
             Univ. Grenoble Alpes, CNRS, IPAG, 38000 Grenoble, France\\
             }

   \date{Received \today}

 
  \abstract{The characterization of exoplanetary atmospheres via transit spectroscopy is based on the comparison between the stellar spectrum filtered through the atmosphere and the unadulterated spectrum from the occulted stellar region. The disk-integrated spectrum is often used as a proxy for the occulted spectrum, yet they differ along the transit chord depending on stellar type and rotational velocity. This is refereed to as the Rossiter-McLaughlin (RM) effect, which is known to bias transmission spectra at high spectral resolution when calculated with the disk-integrated stellar spectrum.
 Recently, it was shown that the first claimed atmospheric signal from an exoplanet cannot arise from absorption in the core of the sodium doublet, because the features observed at high resolution are well reproduced by the RM effect.
  However, it remains unclear as to whether the detection made at medium spectral resolution with the HST arises from the smoothed RM signature or from the wings of the planetary absorption line. More generally, the impact of the RM effect at medium and low spectral resolution remains poorly explored. To address this question, we simulated realistic transmission spectra in a variety of systems using the EVaporating Exoplanets code. We find that the RM effect should not bias broadband atmospheric features, such as hazes or molecular absorption, measured with the JWST/NIRSPEC (prism mode) at low resolution. However, absorption signatures from metastable helium or sodium measured at medium resolution with the JWST/NIRSPEC (G140H mode) or HST/STIS can be biased, especially for planets on misaligned orbits across fast rotators. In contrast, we show that the Na signature originally reported in HD209458b, an aligned system, cannot be explained by the RM effect, supporting a planetary origin. Contamination by the RM effect should therefore be accounted for when interpreting high- and medium-resolution transmission spectra of exoplanets. }


   \maketitle
%
\section{Introduction}

High-resolution ($\mathcal{R} \gtrsim 50000$) spectroscopy has proven to be one of the most prolific techniques for characterizing exoplanetary atmospheres. As a result, spectrographs originally designed to search for exoplanets through velocimetry (such as CARMENES \citealt{Quirrenbach2016}, and ESPRESSO \citealt{Pepe2021}) are increasingly used to probe their atmospheres, from escaping light atoms \citep[see review on helium by][and references therein]{Guilluy2023}, to heavier metals \citep[see e.g.,][]{Hoeijmakers2018,Borsato2023,Prinoth2023}, and molecules \citep{Snellen2010,Birkby2013,Carleo2022}. 
While the narrow absorption lines of these species mainly trace the dynamics and chemistry of the upper atmosphere, medium- and low-resolution instruments provide complementary access to the lower layers through clouds, hazes, and molecules (e.g., \citealt{Lecavelier2008,Snellen2010,Sing2016}).

Atmospheric signatures from transiting planets can be obtained by subtracting the in-transit spectra 
from the out-of-transit stellar spectrum (to isolate the stellar spectrum filtered by the planetary atmosphere), and dividing this difference by a proxy for the occulted stellar lines. Because the emission of the star is not spatially resolved, the disk-integrated stellar spectrum is usually used as a proxy. 

However, the stellar spectrum is not homogeneous across the photosphere and the disk-integrated spectrum is not necessarily representative of local spectra, which vary along the transit chord due to stellar rotation (which induces the Rossiter-McLaughlin (RM) effect \citealt{Rossiter1924,McLaughlin1924}), center-to-limb variations (CLVs) of the line profiles \citep[e.g.,][]{Vernazza1981,Allende2004}, and variation of their continua (broadband limb darkening (LD); \citealt{Knutson2007,Morello2017}) to name but a few. 

Transmission spectra calculated in this way display planet-occulted line distortions (POLDs, \citealt{Dethier2023}) that can bias or even mimic planetary atmospheric signatures. The most famous case is the first claim of an exoplanetary atmosphere through the detection of sodium absorption. \citet{Casasayas2020,Casasayas2021} showed that the signal measured at high resolution in the sodium line mainly arises from POLDs, casting doubt on the planetary origin of the feature measured at medium resolution by \citet{charbonneau_2002_sodium}.

While the biases due to POLDs are now acknowledged and well characterized at high resolution, little is known about their impact at lower resolution, as the poorly resolved stellar lines are then assumed to have negligible effect. Considering the importance of space-based facilities such as JWST \citep{jwst} and Ariel \citep{ariel} for the future of atmospheric characterization, it appears essential that we investigate this question  properly. 

We therefore explored the impact of the RM effect on the transmission spectra of transiting planets measured at low and medium resolution and report our findings in the present paper. 
In Sect.~\ref{sec_transit} we describe the model used to generate synthetic transit spectra. In Sect.~\ref{sec_broad_band_RM} we describe the RM effect in detail, as well as the POLDs that it generates at high resolution, before investigating its impact at low resolution with the JWST. Sect.~\ref{sec_local_RM} focuses on medium-resolution POLDs measured with the HST and JWST in lines of high interest for atmospheric characterization. We discuss our results and conclude in Sect.~\ref{sec_conclusion}.

\section{Generating synthetic stellar spectra}
\label{sec_transit}

\begin{figure}
    \includegraphics[width=\columnwidth]{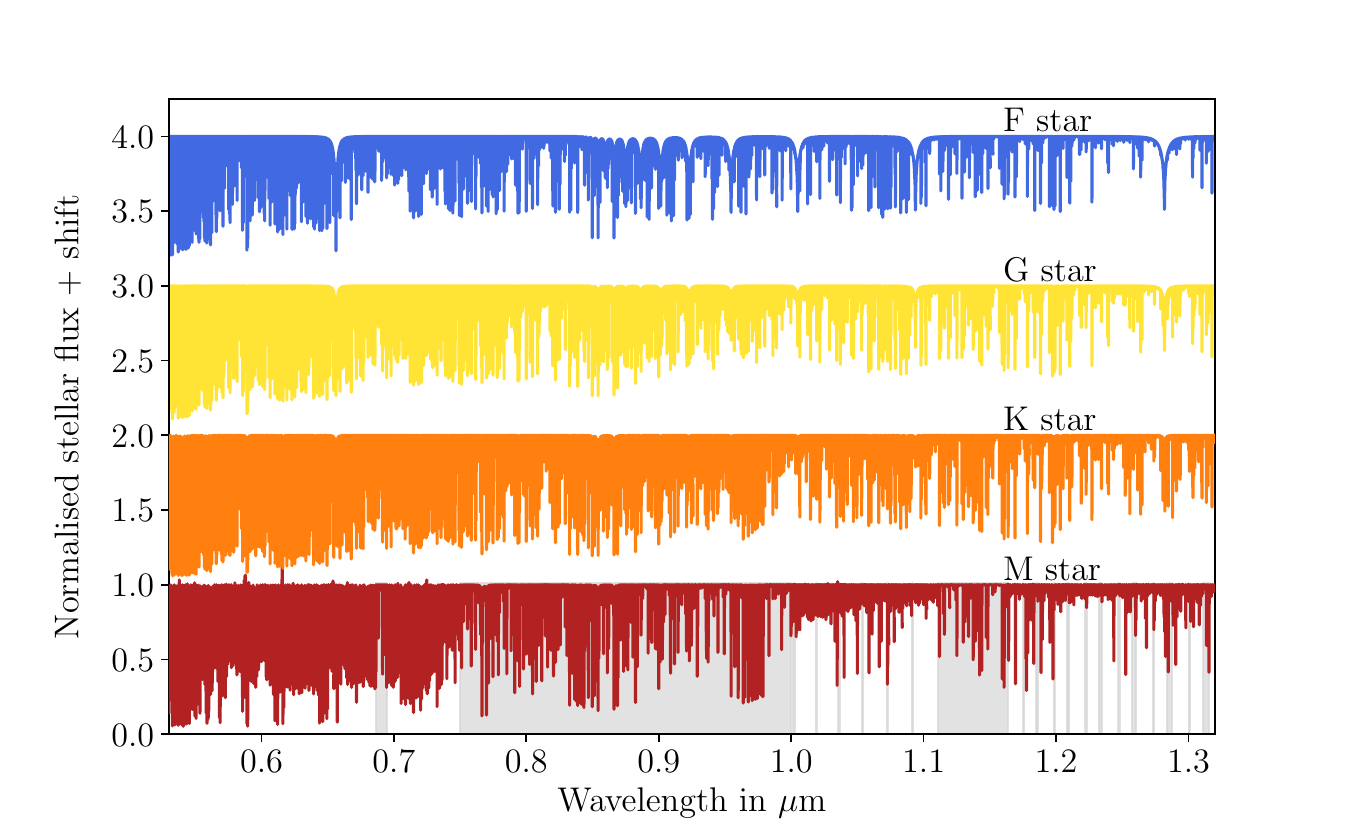}
    \centering
    \caption{Normalized stellar spectra for each spectral type considered (Table \ref{tab_star}). F-, G-, and K-type spectra are generated using \textit{Turbospectrum}. The M-type spectrum combines ESPRESSO and CARMENES observations with \textit{Turbospectrum} in regions too strongly contaminated by tellurics (gray bands).}
    \label{fig_stellar_spec}
\end{figure}

In this section, we describe our procedure to generate synthetic observations of transiting exoplanets with the EVaporating Exoplanets (EVE) code \citep{Bourrier2013,Bourrier2016}. EVE simulates a planetary system in three dimensions, accounting for the associated geometrical effects on transit spectra. The star is modeled as a 2D regular square-cell grid tiled with specific local spectra. This allows spatial variations to be introduced into the stellar spectral emission, and enables the user to account for local occultation of the stellar disk by the opaque continuum of the planet, its extended thermosphere, and escaping exosphere. Based on the occulted stellar grid, EVE generates a time-series of disk-integrated spectra at high temporal cadence and spectral resolution in the chosen spectral bands. The code then convolves these spectra with the adequate instrumental response and resamples them temporally and spectrally, so that they can be compared with observations \citep[see][for more details]{Dethier2023}.

In the present study, we modeled a typical hot Jupiter with the bulk properties of HD\,209458b \citep{Henry2000}. Its transit is simulated across stars of spectral types F, G, K, and M representative of exoplanet hosts, set at the distance of HD\,209458. To maintain the impact parameter and orbital period of the simulated planet at the original values of HD\,209458b, we recomputed the semi-major axis from Kepler's third law and adjusted the orbital inclination accordingly. To assess the net effect of POLDs, we do not simulate an extended atmosphere, but only consider absorption by the opaque regions of the planet. All parameters used in the study can be found in Table~\ref{tab_star}.

To tile the EVE stellar grid, we first generated intensity spectra at the center of the stellar disk using the \textit{Turbospectrum} code\footnote{Latest version, available for download at \href{https://github.com/bertrandplez/Turbospectrum_NLTE/tree/master}{Turbospectrum$\_$NLTE}.} \citep{plez1998,plez2012} and associated line lists (\citealt{heiter2021,magg2022}; VALD3, \citealt{VALD}). We used the interpolation routine\footnote{Developped by Thomas Masseron and Ekaterina Magg.} provided in the \textit{Turbospectrum} package to compute the MARCS photospheric models at the typical temperature, metallicity, and log $g$ of our simulated stellar types (Table~\ref{tab_star}). Because stellar atmosphere models do not accurately reproduce the spectra observed for M dwarf stars \citep{Rackham2023}, the central intensity spectrum of our M dwarf host was approximated using \textit{Turbospectrum} and high-resolution spectra of GJ\,436 measured with ESPRESSO \citep{Bourrier2022} and CARMENES \citep{Nagel2023}. As GJ\,436 rotates slowly (< 1 km\,s$^{-1}$), the observed disk-integrated stellar spectrum can be used as a proxy for the local stellar spectrum at disk center \citep[][neglecting contributions from CLVs and LD]{Dethier2023}. We filled the gaps in the ESPRESSO and CARMENES data with Turbospectrum by joining segments. In each segment, we fit a stellar continuum, and adjust the flux level to ensure continuity of the complete stellar continuum. Spectra are generated between 0.5~$\mu$m (to match JWST/NIRSPEC spectral range) and 1.3~$\mu$m (beyond which there are fewer deep stellar lines). Our final intensity spectra at the center of the stellar disk are shown for each stellar type in Fig.~\ref{fig_stellar_spec}.

We consider the dominant effect of stellar rotation on transmission spectra, neglecting second-order effects due to CLVs \citep{Yan2017,Casasayas2020,Dethier2023}. Stellar rotational velocities are set to typical values for the considered stellar types. Given the large spectral ranges we simulate, we account for broadband LD using a quadratic law \citep{kopal1950} and coefficients computed with the ExoCTK online tool\footnote{Available for download at \href{https://github.com/ExoCTK/exoctk}{ExoCTK}.} \citep{2021zndo...4556063B}. Based on the input rotational velocity and LD properties, the EVE code scales and shifts the intensity spectra in each stellar grid cell to their expected flux level and radial velocity position.

We apply our investigation to realistic synthetic spectra as they would be measured with the HST and JWST. Because the EVE code applies fixed convolution kernels, we generate synthetic spectra in independent, consecutive bands over which the true instrumental resolving power varies little. The time-series of synthetic spectra are convolved with a Gaussian profile, the width of which is set by the average resolving power $\mathcal{R}$ over each spectral band. Convolved spectra are then resampled over the instrument wavelength table. Finally, transmission spectra $\mathcal{T}$ are calculated as the ratio between each in-transit spectrum $f_\mathrm{in}$ and the simulated disk-integrated spectrum $f_\mathrm{\star}$, processed in the same way:
\begin{equation}
    \mathcal{T}(\lambda, t, \mathcal{R}) = \frac{f_\mathrm{in}(\lambda, t, \mathcal{R})}{f_\mathrm{\star}(\lambda, \mathcal{R})},
    \label{eq_transmission}
\end{equation}
where $t$ is the time relative to mid-transit and $\lambda$ is the wavelength. To isolate the net contribution of the POLDs, we define differential transmission spectra by subtracting the constant continuum from the opaque regions of the planet. The opaque continuum of the planet is calculated by simulating its transit across non-rotating stars accounting for LD.

\newcolumntype{P}[1]{>{\centering\arraybackslash}p{#1}}
\setlength{\tabcolsep}{2pt}
\begin{table}[]
\caption{Stellar and planetary parameters used in this work.}
\centering
\hspace{-0.25cm}
\begin{tabular}{P{2.6cm}|P{1.3cm}|P{1.66cm}|P{1.65cm}|P{1.03cm}|}

\multicolumn{5}{c}{\textit{Stellar parameters}}                                                                                                               \\ 
\cmidrule{1-5}\morecmidrules\cmidrule{1-5}

\\[-1em]
Name                         & HR8799$^{a}$ & HD209458$^{b}$ & HD189733$^{c}$ & GJ436$^{d}$                                \\
Stellar type                            & F     & G        & K        & M                                        \\  
T$_{\mathrm{eff}}$ [K]         & 7400       & 6065         & 5050  & 3500             \\
Radius [R$_{\odot}$]    & 1.5       & 1.15         & 0.77  &  0.42          \\
Mass [M$_{\odot}$]     & 1.60       & 1.16         & 0.83   &  0.44         \\
log $g$ [cm\,s$^{-2}$] & 4.20       & 4.36         & 4.58  &   4.84                \\
v$_{eq} \sin(i)$ [km\,s$^{-1}$] & 20       & 10$^{\dagger}$         & 5     & 0.5                                     
\\ \multicolumn{5}{c}{} \\[-0.5em]
\multicolumn{5}{c}{\textit{Planetary parameters$^e$}}                                                                                                               \\ \cmidrule{1-5}\morecmidrules\cmidrule{1-5}
\\[-1em]
Radius [R$_{\mathrm{jup}}$]                       & \multicolumn{4}{c|}{1.3957}                             \\
Mass [M$_{\mathrm{jup}}$]                      & \multicolumn{4}{c|}{0.68}                                 \\
Period [day]                      & \multicolumn{4}{c|}{3.52475}                                 \\
Spin-orbit angle $\lambdaup$ [$^\circ$]                      & \multicolumn{4}{c
|}{}                                 \\[-1.75em]
                     & \multicolumn{4}{c|}{1.58$^{\ddagger}$}                                \\[0.57em]

Semi-major axis [AU]  & & & &   \\[-1.75em]    

 & 0.053       & 0.047         & 0.043  & 0.034             \\[0.57em]
Orbital inclination [$^\circ$]        & & & &   \\[-1.75em]
& 86.29       & 86.78         & 87.60  & 88.42              
\\[0.57em] \multicolumn{5}{c}{} \\[-0.5em]
\end{tabular}
{\raggedright \small Stars are set at 47\,pc distance, with solar abundances, and rotational velocities typical of the chosen type. $\dagger$: v$_{eq} \sin(i)$ is set to the \citet{Casasayas2021} value (4.23 km\,s$^{-1}$) for HD\,209458 in Sec.~\ref{Sec_Na_line}. $\ddagger$: $\lambda$ is set to different values (from -90 to 90) in Secs.~\ref{sec_broad_band_RM}~\&~\ref{Sec_He_line}. $a$:~\citet{Stassun2019}. $b$:~\citet{Torres2008}. $c$:~\citet{Addison2019}. $d$:~\citet{Bourrier2022,Nagel2023}. $e$:~\citet{Casasayas2021}. \par}
\label{tab_star}
\end{table}


\section{Broadband RM effect}
\label{sec_broad_band_RM}

\begin{figure*}
    \includegraphics[width=2\columnwidth]{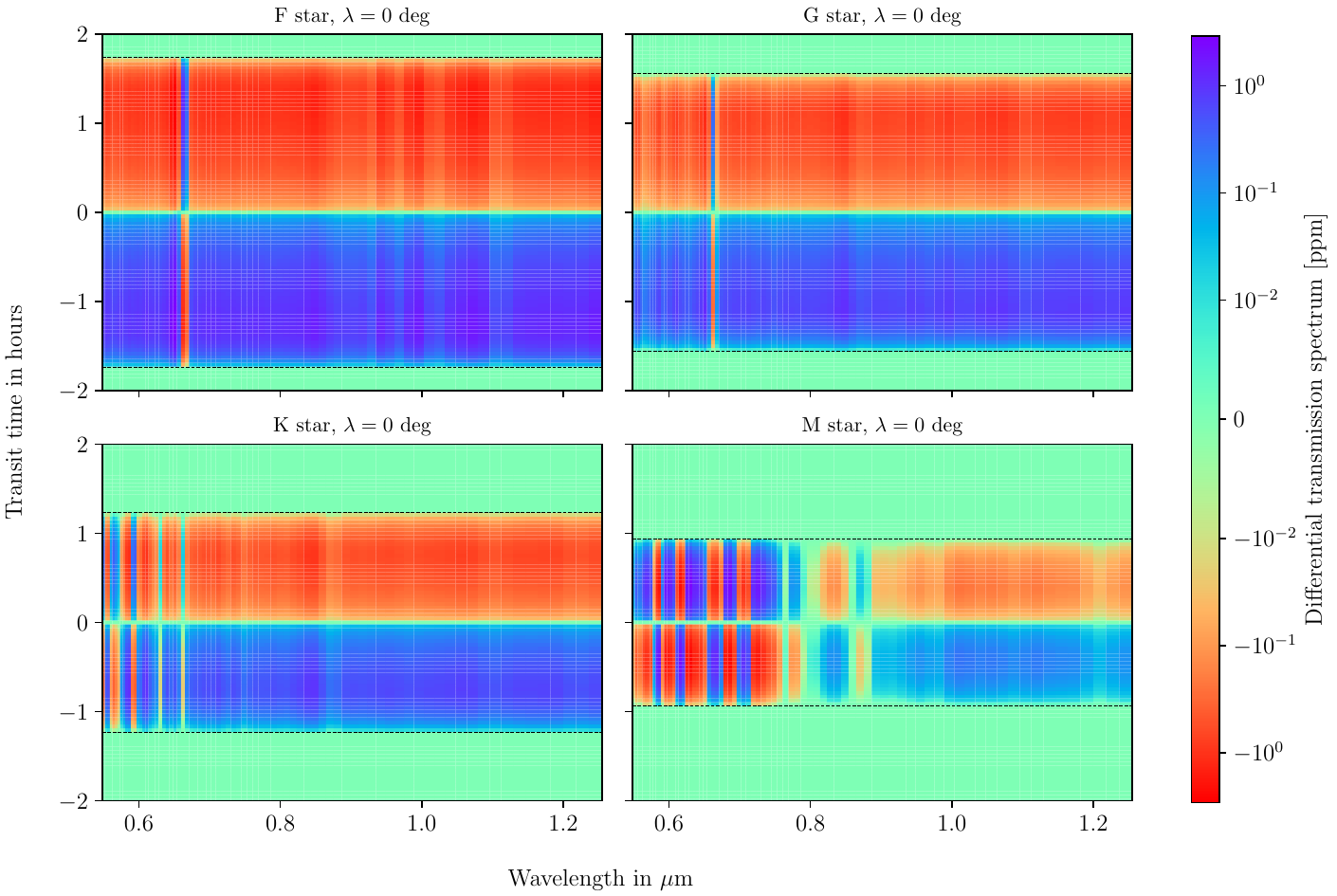}
    \centering
    \caption{Synthetic time-series of differential transmission spectra for our four test hosts and an aligned orbit, as observed with JWST/NIRSPEC in prism mode ($30 \le \mathcal{R} \le 65$). Dashed black lines represent the beginning and end of the transits.}
    \label{fig_map_stellar}
\end{figure*}

In this section, we investigate the extent to which the RM effect distorts transmission spectra measured over large wavelength bands and at low resolution ($\mathcal{R}\sim100$). We expect the RM effect and the resulting POLDs to depend on the properties of the planet-occulted regions, controlled at first order by the stellar rotational velocity v$_{eq} \sin(i)$ and sky-projected spin-orbit angle $\lambdaup$ \citep[see][and reference therein]{Albrecht2022}. Indeed, as a planet transits, it occults regions of the stellar photosphere moving with different radial velocities, and thus emitting spectra at different Doppler shifts. At high spectral resolution, the RM-induced POLDs further depend on the shape of the occulted stellar line profiles. However, whether or not those various properties have the same influence on the POLDs measured at low resolution is unclear.

Because of its interest for atmospheric characterization, we generated synthetic spectra for the JWST instrument NIRSPEC in prism mode, which can observe from 0.6 to 5.3~$\mu$m. We limited our simulations to the wavelength range with the highest number of lines, namely $0.55-1.25$~$\mu$m. We extracted the wavelength table from the JWST Exposure Time Calculator\footnote{Available online at \href{https://jwst.etc.stsci.edu/}{JWST ETC}.} (ETC) and extended this range up to 0.5 microns using the tables in \citet{Rustamkulov2023}.
We divided stellar spectra into independent bands of 0.1~$\mu$m, processed with specific resolving power\footnote{From the tables in the \href{https://jwst-docs.stsci.edu/jwst-near-infrared-spectrograph/nirspec-instrumentation}{JWST documentation}.} and LD coefficients.

\subsection{Impact of the stellar spectrum profile}

\begin{figure*}
    \includegraphics[width=2\columnwidth]{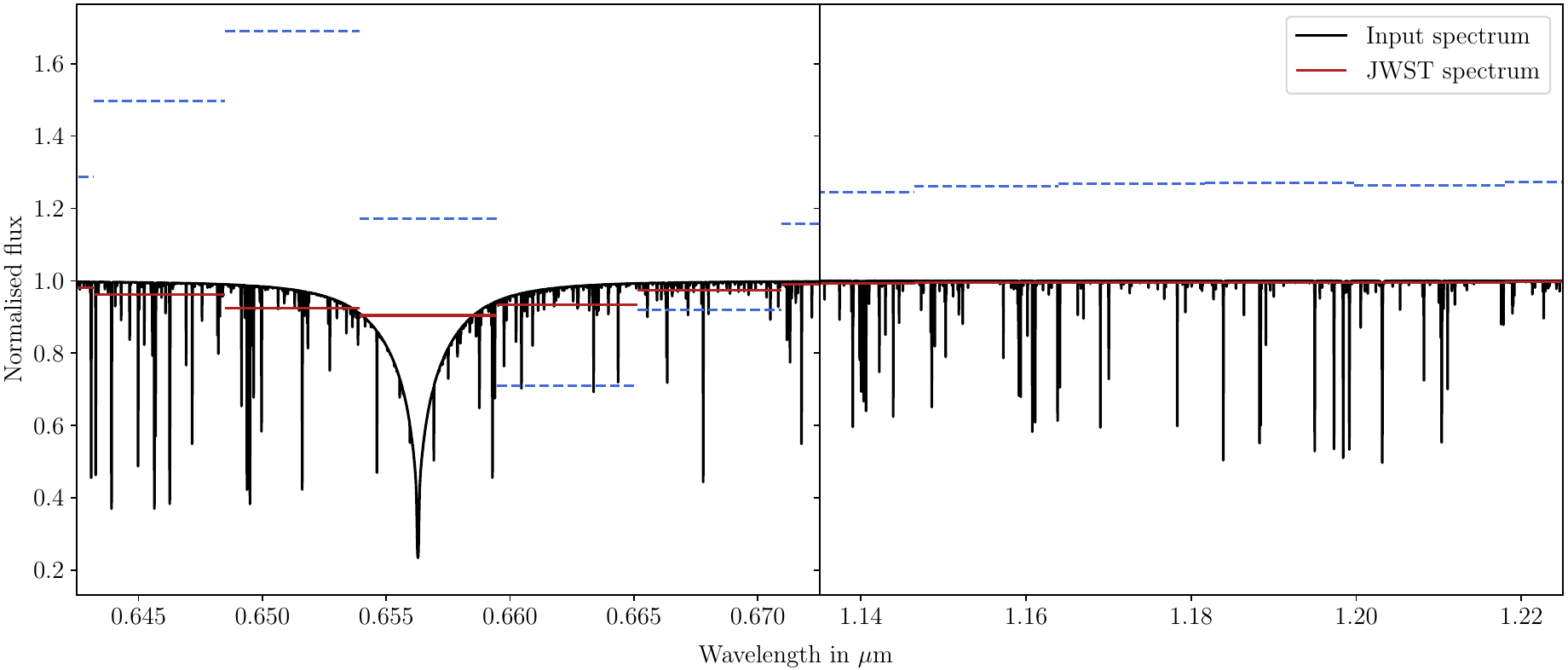}
    \centering
    \caption{Normalized stellar spectrum of the F-type star at high resolution (black) and as seen by JWST/NIRSPEC (red bins) in prism mode ($30 \le \mathcal{R} \le 65$). Dashed blue bins represent the differential transmission spectrum at $T=-0.25$ hours (see Fig.~\ref{fig_map_stellar}) arbitrarily scaled. The differential transmission spectrum was shifted by 1 to be visible together with the flux spectrum. The left panel shows a zoomed-in image of a deep and broad stellar line. }
    \label{fig_jwst_spec}
\end{figure*}

Figure~\ref{fig_map_stellar} shows the difference maps between the transmission spectra (Eq.~\ref{eq_transmission}) with stellar rotation and without, which corresponds to the differential transmission spectra defined in Sect.~\ref{sec_transit}. In particular, we compare different stellar types in the case of an aligned orbit ($\lambdaup=0^\circ$). Going from F- to M-type, we see increasing numbers of S-shaped features (inversion of sign between consecutive bins) appearing at short wavelengths in the differential transmission spectra.
We think that this is due to the high density of deep and broad absorption lines in the stellar spectrum at these wavelengths (see Fig.~\ref{fig_stellar_spec}), which increases as stars get colder. To support this idea, we show in Fig.~\ref{fig_jwst_spec} the F-type stellar spectrum as seen by JWST/NIRSPEC. We observe stellar lines that are wide enough to cover more than one instrumental pixel, and that can still be distinguished after instrumental convolution in the low-resolution stellar spectrum. The RM effect works in the same way for these smoothed stellar lines as at high resolution, creating an "S" shape around the central pixel (blue bins in Fig.~\ref{fig_jwst_spec}). At higher wavelengths of $\gtrsim 0.8$ microns, pixels are larger and the resolving power smaller, which means that the smoothing is performed over a larger number of stellar lines, mostly shallow and narrow, which are not spectrally resolved and do not generate sharp features (see right panel in Fig.~\ref{fig_jwst_spec}). The resulting pixel-to-pixel variations are therefore smaller in the differential transmission spectra, even though they are still visible around the widest stellar lines. 
Furthermore, all spectral types show an additional bias manifesting as an offset over the full range of the differential spectra, changing sign at mid-transit. When a planet occults a redshifted (blueshifted) local line, the in-transit disk-integrated line is distorted in such a way that its profile appears blueshifted (redshifted)  overall. This is why the RM effect translates into an anomalous shift in the RV centroid of the disk-integrated stellar lines. Here, we attribute the bias to a similar effect arising from the stellar continuum. The ratio of the in-transit to out-of-transit spectral continua is smaller (larger) than the expected planetary continuum for a stellar continuum with a negative (positive) slope, and therefore yields a negative (positive) offset in the differential transmission spectrum. The offset changes sign at some point during the transit (here at mid-transit because of the aligned orbit) when the planet crosses the projected stellar spin axis and the Doppler shift of the occulted local spectra changes sign. Another way to understand this bias is by considering the stellar continuum as the wing of a very shallow and broad stellar line, upon which the RM effect acts as usual. This bias only appears at low spectral resolution, as the RM induced by the numerous narrow stellar lines approximately averages to zero and makes the RM effect of the stellar continuum  visible. 
We note that, because of strong tellurics, between 0.8 and 1.0 $\mu$m, the M-type spectrum had to be reconstructed using \textit{Turbospectrum}. The stellar continuum is almost flat in these wavelengths, leading to smaller biases in the differential transmission map. At low wavelengths, the M-type stellar continuum slope switches sign several times. Together with several broad and deep lines in this spectral region, this behavior explains the more complex spectral variations of the POLDs observed for the M dwarf host.

\subsection{Impact of stellar rotation}

The rotational velocity of the host star does not significantly modify the spectro-temporal structure of the maps. This is because the instrumental bins and convolution kernel are much larger than the Doppler shift induced by the stellar rotational velocity. The change in spectral distribution of the planet-occulted stellar lines therefore has a negligible impact on the in-transit disk-integrated spectra measured with NIRSPEC at low resolution, and an increase in v$_{eq} \sin(i)$ only increases the magnitude of the POLDs (similarly to RM-induced POLDs at high-resolution; \citealt{Dethier2023}).

\subsection{Impact of transit chord location}

\begin{figure}
    \includegraphics[width=\columnwidth]{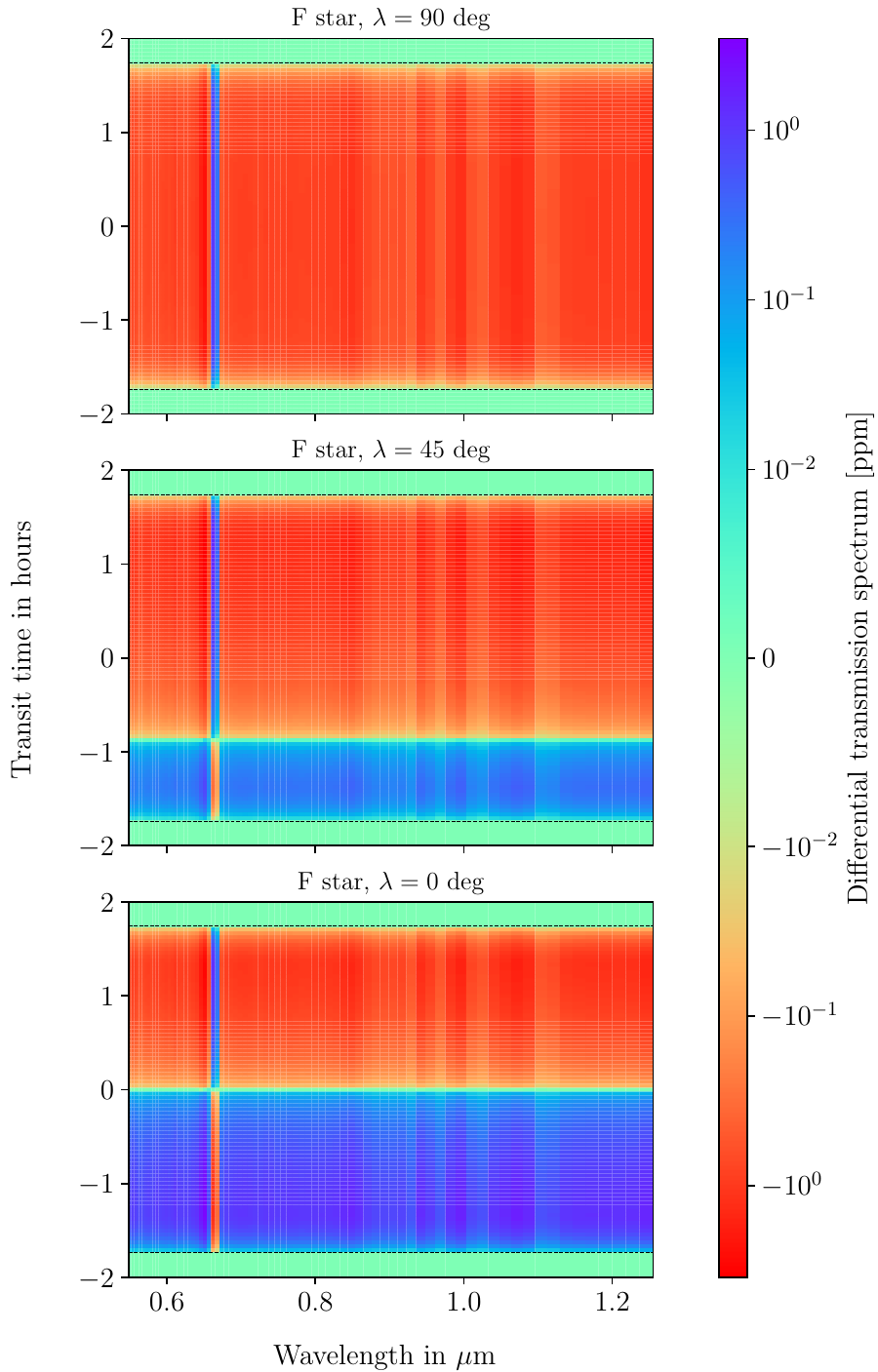}
    \centering
    \caption{Synthetic time-series of differential transmission spectra for the F-type host and different spin--orbit angles, as observed with JWST/NIRSPEC in prism mode ($30 \le \mathcal{R} \le 65$). Dashed black lines represent the beginning and end of the transits.}
    \label{fig_map_lambda}
\end{figure}

We fixed the stellar spectrum to the F-type star and compared the change of the RM bias with respect to $\lambdaup$ during the full transit. In Fig.~\ref{fig_map_lambda} we directly notice that the maps show a very different time evolution. The instrumental response at low resolution is so broad that it covers several stellar lines. As a result, the RM-induced POLDs are merged and generate structure with a more global behavior than at high spectral resolution. 
The value of the spin--orbit angle mainly determines the time during transit at which the planet crosses the sky-projected stellar spin, and thus at which the low-resolution POLDs and continuum offset change sign. For an aligned orbit ($\lambdaup=0^\circ$), these features therefore reverse exactly at the center of the transit because the planet occults  local spectra that are blueshifted during the first half of the transit, and redshifted for the second half. In the case of a polar orbit ($\lambdaup = 90^\circ$), the planet moves parallel to the axis of stellar rotation and occults local spectra with constant Doppler shifts, so that features in the differential transmission spectra remain of the same sign. For the intermediate cases (e.g., $\lambdaup = 45^\circ$), the moment when the differential transmission spectra changes sign is determined by the exact geometry of the orbit.

\subsection{Consequences for planetary signatures}

The RM-induced biases we observe in our simulations could have a significant impact on atmospheric characterization at low spectral resolution, depending on their detectability. Using Pandexo\footnote{Available online at \href{https://exoctk.stsci.edu/pandexo/calculation/new}{Pandexo}.} \citep{2017PASP..129f4501B}, we estimated the errors that would be associated with observations of our four systems. We considered favorable cases with a long observation baseline (so that errors on the master-out stellar spectrum are negligible), $\lambdaup=90^\circ$ to maximize the average POLDs over the transit, and no systematic errors. This yields uncertainties on the differential transmission between of $\sim$120 and 40 ppm, which are much larger than the maximum amplitude of a few parts per million (ppm) reached by the simulated POLDs (for the $\lambdaup=90^\circ$ case). Furthermore, the predicted errors are largest between 0.55 and 0.7 $\mu$m, where the POLDs are strongest. For completeness, we generated a basic transmission spectrum from a planetary atmosphere composed of hydrogen and helium using \textit{Exo\_k} \citep{Leconte2021}, and added it to the differential transmission spectrum for the $\lambdaup=90^\circ$ case\footnote{This is not accurate, as the planet does not absorb directly the disk-integrated stellar spectrum, but it is enough for the purpose of our test.}. We fitted the resulting transmission spectrum with the same atmospheric model, retrieving the input parameters with success. We therefore conclude from our investigations that the RM effect is unlikely to bias planetary transmission spectra measured at low spectral resolution, even in the most extreme cases.

\section{RM effect from isolated stellar lines}
\label{sec_local_RM}

In this section, we investigate the impact of the RM effect on transmission spectra measured at medium resolution ($\mathcal{R} \sim$ a few thousands), focusing on two transitions of great interest for exoplanet characterization: the sodium doublet ($\lambda \sim 5890 \text{\AA}$) and the metastable helium triplet ($\lambda \sim 10830 \text{\AA}$).

\subsection{Sodium doublet}
\label{Sec_Na_line}

\begin{figure*}
    \includegraphics[width=2\columnwidth]{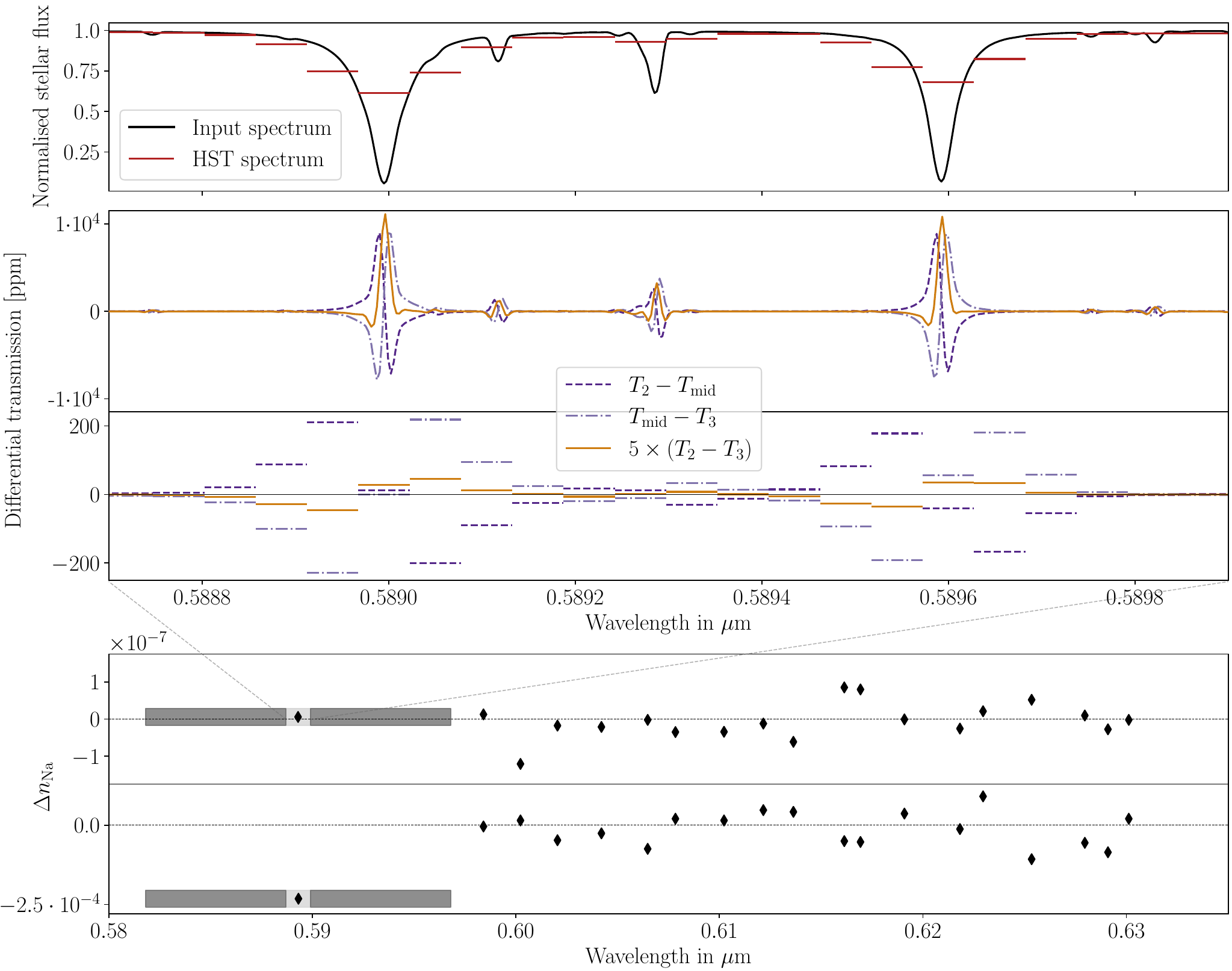}
    \centering
    \caption{Sodium analysis of HD\,209458b. All spectra are shown in the stellar rest frame. \textit{First panel (top)}: Synthetic spectrum of HD\,209458 used in our simulations (black, at high resolution) and as measured with HST/STIS (red, 0.55\,\AA\,-wide pixels, $\mathcal{R} = 5440$). \textit{Second panel}: Simulated differential transmission spectra of HD\,209458b without atmosphere as measured with ESPRESSO (0.04\,\AA\,-wide pixels, $\mathcal{R} = 140000$), averaged over different phases of the transit. \textit{Third panel}: Simulated differential transmission spectrum of HD\,209458b as measured with STIS and averaged over the full transit. \textit{Fourth panel}: Simulated $\Delta n$ (Eq.~\ref{Eq_delta_na}), comparable to the observed values in the fifth panel. All points are computed using the same three bins defined in Eq.~\ref{Eq_bins}, centered on different stellar lines. Shaded areas are associated with the sodium doublet (light gray for the central bin $b_c$, dark gray for the red and blue bins $b_r$ and $b_b$). \textit{Fifth panel (bottom)}: Measured $\Delta n$ from Fig.~5 in \citet{charbonneau_2002_sodium}.}
    \label{fig_na}
\end{figure*}

The first detection of an exoplanet atmosphere was claimed through the detection of sodium absorption in medium-resolution ($\mathcal{R} \sim 5440$) HST data for HD\,209458b \citep{charbonneau_2002_sodium}. This result was supported by independent reanalyses of the same HST dataset \citep{Sing2008,Vidal2011,Morello2022}, and new observations at high resolution of the sodium doublet (e.g., \citealt{Snellen2008} with Subaru/DRS or \citealt{Langland2009} with HIRES/KECK). However, this detection was recently put into doubt \citep[see][]{Casasayas2020,Casasayas2021,Dethier2023} by the realization that the sodium absorption signal measured with ESPRESSO at high spectral resolution ($\mathcal{R} \sim 140000$) can be reproduced by POLDs induced by the RM and CLV effects. It is currently unclear whether the signature observed at medium resolution could still arise from the planetary atmosphere or result from the same stellar contamination as observed at high resolution \citep{Morello2022}.

We simulated synthetic spectra from HD\,209458b as observed by \citet{charbonneau_2002_sodium} using HST/STIS in the G750M mode. As our goal is to assess whether or not RM-induced POLDs can reproduce the sodium signature measured with STIS, we do not include an atmosphere in our model planet. The rotational velocity of the model G-type star is set to 4.23 km\,s$^{-1}$, the value for HD\,209458. Wavelength tables are defined using the HST ETC\footnote{Available online at \href{https://etc.stsci.edu/}{HST ETC}.}. Figure~\ref{fig_na} summarizes the steps of our analysis.

We first calculated differential transmission spectra at the resolution of ESPRESSO (Fig.~\ref{fig_na}, second panel). When averaging these spectra over the full transit in the planet rest frame (not shown here), we obtain POLDs similar to those found by \citet{Casasayas2021}, thus validating our procedure. We then averaged the differential spectra in the stellar rest frame over different phases of the transit. Due to the near-aligned orbit ($\lambdaup = 1.58^\circ$), the planet occults regions at $-t$ and $t$ whose local spectra have the same profile but opposite Doppler-shift, meaning that the POLDs over $T_2$--$T_\mathrm{mid}$ and $T_\mathrm{mid}$--$T_3$\footnote{$T_{2/3/\mathrm{mid}}$ are respectively the second/third contact and mid-transit times.} have opposite S-shape profiles. When averaged between $T_2$ and $T_3$, these features ---which are nearly symmetrical with respect to the line transition (\textit{i.e.}, to a null RV at the intersection of the transit chord with the projected stellar spin axis at the transit mid-point)--- result in a POLD with a W-shaped profile of smaller amplitude.

We then calculated differential transmission spectra at the medium-resolution and sampling of HST/STIS (Fig.~\ref{fig_na}, third panel), averaging over different phases of the transit in the stellar rest frame. Because the sodium lines are sampled by several instrumental pixels, they remain visible in the medium-resolution stellar spectrum and still induce a POLD. We still observe "S" shapes for the two halves of the transit, albeit with a smaller amplitude than at high resolution because the convolved stellar lines are shallower. However, when averaged over the full transit, the amplitude of the POLD is reduced by about two orders of magnitude compared to high resolution, down to only a few ppm; it loses its W shape because it is spectrally narrower than one HST pixel of HST, and thus cannot be resolved. 
At the medium resolution of the HST, the shape of the transit-average POLD becomes dominated by the slight misalignment of the orbit. The absorption of redshifted local stellar spectra is slightly longer and contributes more to the average differential transmission spectrum.

In the last two panels of Fig.~\ref{fig_na}, we compare our results with those of \citet{charbonneau_2002_sodium}, using their procedure to analyze the sodium doublet. \citet{charbonneau_2002_sodium} obtained time series of spectra from HD\,209458 during four transits with HST/STIS. 
After calculating a master out-of-transit spectrum $f_{\mathrm{out}}$ in each visit, these authors defined transmission spectra (see Eq.~\ref{eq_transmission}) for each exposure, and then integrated $\mathcal{T}(\lambda, t)$ in three bins: 
\begin{eqnarray}
\label{Eq_bins}
    && b_{b}  \in  \mathrm{[581.8-588.7] nm}, \nonumber \\
    && b_{c} \in  \mathrm{[588.7-589.9] nm},   \\
    && b_{r} \in  \mathrm{[589.9-596.8] nm}, \nonumber
\end{eqnarray}
yielding time series $n_{b,c,r}(t)$. The goal of these authors was to isolate a possible planetary sodium feature in $n_{c}$ by removing the contribution from broadband atmospheric and stellar variations measured in the blue and red bins ($b_b$, $b_r$). To do so, they defined the quantity of interest 
\begin{equation}
    \label{Eq_delta_na}
    \Delta n_{\mathrm{Na}} = n_c - \frac{1}{2} (n_b+n_r),
\end{equation}
which they averaged over the full transit. They further computed $\Delta n$ at 18 other wavelengths using the same three bins centered on other stellar lines.

Panel 4 of Fig.~\ref{fig_na} compares the $\Delta n$ series measured by \citet{charbonneau_2002_sodium} and the one derived from our simulations. The observed values show a clear signal in absorption (4.1 $\sigma$) at the position of the sodium doublet (\textit{i.e.,} in $b_{c}$), while the flux ratio is stable in the other spectral lines. Our simulated values, which only vary because of the POLDs induced by the RM effect of a planet with no atmosphere, are stable in all lines. In particular, the simulated $\Delta n_{\mathrm{Na}}$ is $\sim 10^4$ times smaller than the observed value. This is because the central bin $b_c$ covers about 22 HST pixels, encompassing the full range of the sodium doublet. As can be seen in Panel 3 of Fig.~\ref{fig_na}, the medium-resolution POLDs averaged over the full transit display roughly symmetrical "S" shapes around each line transition within $b_c$, meaning that their integrals cancel out in $\Delta n_{\mathrm{Na}}$. Our analysis therefore shows that the RM effect, while generating strong absorption- and emission-like features at high resolution in the core of the sodium lines, cannot explain the broadband absorption feature as measured by \citet{charbonneau_2002_sodium}.

 \citet{Morello2022} proposed that the sodium signal detected by \citet{charbonneau_2002_sodium} could be caused by limb-darkening contamination. However, since the measured $\Delta n$ are stable in other spectral lines (RMS value of $5.3\cdot 10^{-5}$, which yields a ratio $\Delta n_\mathrm{Na}$/RMS $\simeq 4.4$ ) this hypothesis appears unlikely. An alternative possibility is that the broadband sodium signal does arise from the planetary atmosphere, more particularly from the wings of the sodium absorption profile, given that the high-resolution signal in the core is not of planetary origin. Such a scenario could be explained by the presence of sodium in the troposphere but not in the thermosphere (e.g., \citealt{Pino2018}). Tropospheric sodium would need to be visible in absorption to generate the observed wings, imposing low temperatures and/or aerosols at low altitudes \citep[e.g.,][]{Morello2022}. Nevertheless, thermospheric sodium could be absent because of condensation or ionization (e.g., \citealt{Vidal2011}).

\subsection{Helium triplet}
\label{Sec_He_line}

\begin{figure*}
    \includegraphics[width=2\columnwidth]{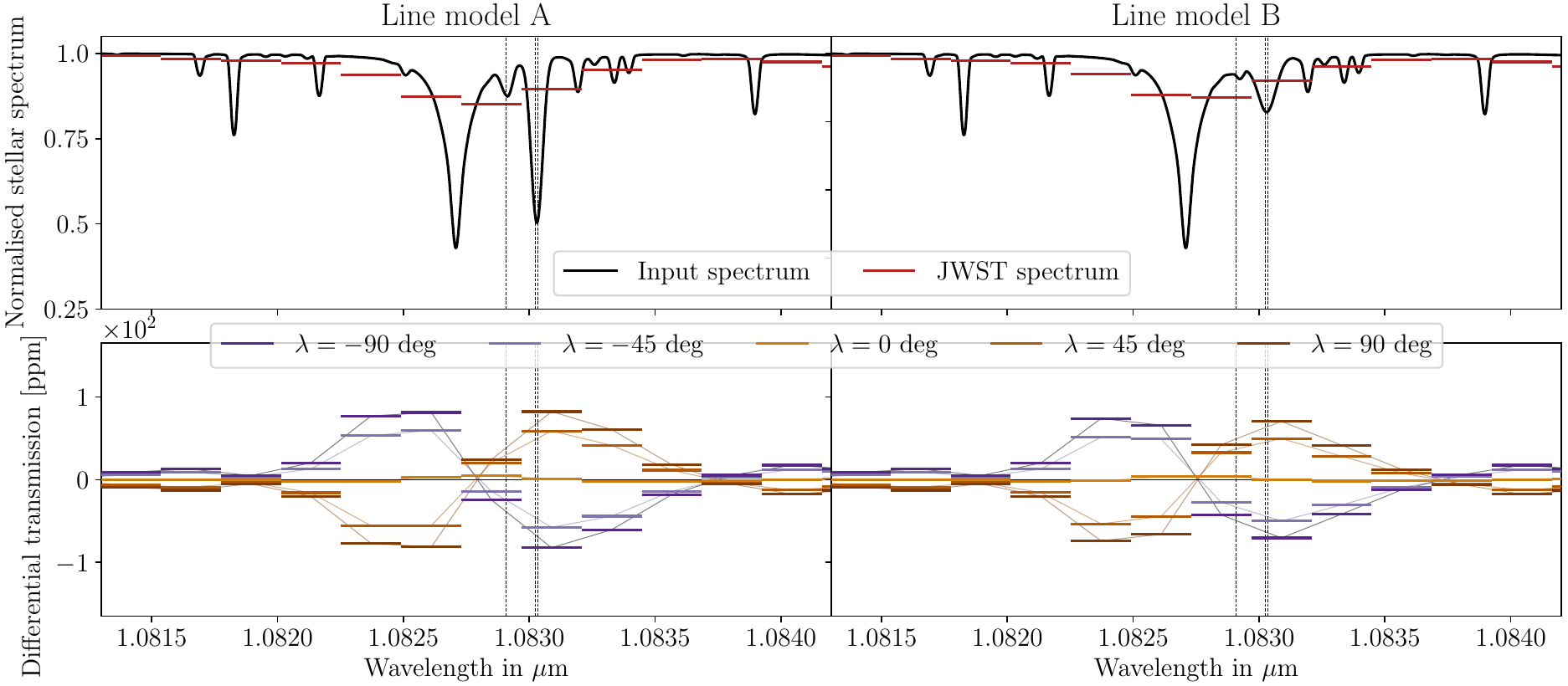}
    \centering
    \caption{Helium analysis. Black dashed lines highlight the three metastable lines. \textit{Top panel}: Synthetic spectrum of the K-type star used in our simulations (in black, at high resolution) and as measured with JWST/NIRSPEC in G140H-f70 mode (in red, 2.4\,\AA\,-wide pixels, $\mathcal{R} = 2050$) in the star rest frame. \textit{Bottom panel}: Simulated differential transmission spectra as measured with NIRSPEC in the planet rest frame, averaged over the full transit and shown for different spin--orbit angles. }
    \label{fig_he}
\end{figure*}

Over recent years, extended thermospheres have been unveiled through high-resolution spectroscopy in the metastable helium triplet \citep[see e.g.,][]{Oklopcic2018,Guilluy2023}. Several studies already discussed the possibility of using this tracer at lower resolution but higher sensitivity with the JWST in order to characterize atmospheric escape from smaller planets \citep{Allart2018,DosSantos2023}. This again raises the question of whether or not the RM effect can bias the detection of planetary helium absorption at medium resolution. 

We simulated synthetic transmission spectra around the helium triplet, as observed with the JWST/NIRSPEC in G140H-f70 mode ($\mathcal{R}\sim 2050$). We considered the case of the K-type star (Table \ref{tab_star}), thought to enhance the population of metastable helium level in exoplanet atmospheres \citep{Oklopcic2019}. The MARCS models that we use as inputs for \textit{Turbospectrum} only model the stellar photosphere. However, the conditions required for the helium triplet metastable state to be populated are usually met in the regions of the stellar atmosphere above the photosphere, namely the chromosphere and transition region \citep{Andretta1997}. Therefore, the helium triplet lines are not simulated in our synthetic stellar spectra. Nevertheless, we included these lines by multiplying the intensity spectrum at disk center (derived with \textit{Turbospectrum}) with an analytical Gaussian line profile using the properties of the helium atom with an arbitrary density and temperature (see \citealt{Guilluy2023}). To generalize our investigation, we take two extreme cases for the stellar helium lines (top panels of Fig.~\ref{fig_he}), either narrow and deep ($\sim 71\%$ of the silicon line; model A) or wide and shallow ($\sim 25\%$ of the silicon line; model B). The lower panels of Fig.~\ref{fig_he} display differential transmission spectra averaged in the planetary rest frame over the $T_2-T_3$ window for the two line models and various transit chords.

The deep stellar silicon line is spectrally close enough to the helium triplet that it dominates the stellar spectrum at the resolution of JWST/NIRSPEC, and is the main contributor to the RM-induced POLD. Moreover, the width of the silicon line is comparable to the size of instrumental pixels. Similarly to the sodium case (Sect.~\ref{Sec_Na_line}), the resulting medium-resolution POLDs form an "S" shape. The comparison between cases A and B shows that the blended metastable helium lines only induce second-order variations in the POLD, with a stronger amplitude for a deeper helium triplet. We also generated differential transmission spectra for different values of the spin--orbit angle. As expected from high-resolution studies (e.g., \citealt{Dethier2023}), the RM-induced POLDs vary strongly in shape with $\lambdaup$. As such, aligned orbits ($\lambdaup=0^\circ$) produce POLDs that are more than an order of magnitude smaller than those produced by polar orbits ($\lambdaup=\pm 90^\circ$).

To perform realistic predictions, we used Pandexo to compute  the uncertainties on the stellar spectrum in a given exposure, and propagated them on the averaged differential transmission spectra. Assuming a reasonable baseline of 5 hours, we estimated errors of the order of 100 ppm in the helium triplet, which is comparable to the amplitude of the simulated POLDs for misaligned orbits. Even if the POLD itself is smaller in amplitude than the atmospheric helium signal expected from evaporating planets (on the order of a few tenths of a percent at medium resolution), this means that the RM effect can bias such signatures even at the medium resolution of the JWST/NIRSPEC. While planets on aligned orbits should remain little affected because of the smoothing of the POLD when averaged over the full transit, the bias would be strongest for long transits/bright stars (reducing flux errors) and fast rotators/polar orbits (increasing POLD amplitude).  

We therefore recommend that the amplitude of RM-induced POLDs be evaluated before modeling atmospheric signatures in JWST transmission spectra, as is already done at high resolution. If contamination is expected, POLDs should be accounted for in the model, ideally using global simulations of the stellar and planetary absorption lines, as done with the EVE code (\citealt{Dethier2023}).

\section{Discussion and conclusion}
\label{sec_conclusion}

In this study, we evaluated the impact of the RM effect on exoplanet transmission spectra measured at low and medium spectral resolution. To do so, we used a 3D forward model, the EVE code, to simulate synthetic transmission spectra from a typical hot Jupiter, HD\,209458b, transiting across realistic stellar grids representative of F, G, K, and M stars.

Before discussing our results on these systems, we want to highlight the problematic nature of comparing observed and synthetic transmission spectra. It is common in the literature to directly convolve and resample theoretical transmission spectra. However, to reproduce observations accurately, flux spectra must first be convolved and resampled before transmission spectra are computed. We estimate that the two methods only yield similar results if the FWHM of the spectrograph LSF is not too large ($\mathcal{R} < 1000$) or too close to the FWHM of the stellar lines. However, we caution that POLDs of small amplitude may be affected by an improper convolution regardless of the width of the instrumental response.

To evaluate biases at low resolution, we simulated spectra as measured with the JWST/NIRSPEC in prism mode, over 0.55-1.25~$\mu$m. In contrast to high-resolution spectroscopy, where the RM effect induces isolated POLDs at the spectral position of planet-occulted stellar lines, we found that the broad instrumental convolution and resampling blend the POLDs and creates structured noise along the transit. Besides the expected dependency of the noise structure on the spin-orbit angle, we note a strong influence of stellar type through the spectral distribution of stellar lines. The strongest bias in transit-averaged transmission spectra is observed for polar orbits and late-type stars, whose deep and numerous lines induce strong spectral variations between 0.55 and 0.8 $\mu$m. Nonetheless, we find that the amplitude of the POLDs remains small enough in all cases that they will have no impact on planetary atmospheric measurements performed with the JWST/NIRSPEC at low resolution.

We also simulated spectra at medium resolution to evaluate biases in specific lines tracing the upper atmosphere of exoplanets:
\begin{itemize}
    \item Helium in the thermosphere: POLDs overlapping with the planetary helium lines are induced by the blended stellar silicon and helium lines. The spectral resolution of the JWST/NIRSPEC (G140H mode) allows transmission spectra to be averaged in the planet rest frame in order to enhance atmospheric signatures. The resulting average POLD is smoothed out and induces no bias in aligned systems but is amplified to the level of uncertainties on the absorption signal in misaligned systems. RM-induced POLDs should therefore be accounted for when interpreting helium transmission spectra of such systems, especially for long transits across bright and fast-rotating stars. 
    \item Sodium in the troposphere: we revisit the apparent discrepancy in HD\,209458b between the absorption feature measured at medium resolution with HST/STIS ---claimed as the first exoplanet atmospheric detection (\citealt{charbonneau_2002_sodium})--- and the absorption feature measured at high resolution with ESPRESSO, which was explained as POLDs in the core of the stellar sodium lines. Processing our synthetic STIS spectra of HD\,209458b with the procedure of \citealt{charbonneau_2002_sodium}, we obtained POLDs on the order of 0.01\,ppm in the sodium doublet, which is negligible compared to the measured signal of 250\,ppm. Our results strengthen the hypothesis that the medium-resolution sodium signal of HD\,209458b traces the wings of its absorption line, arising from atoms above the opaque layers but removed by an unknown process at high altitudes. 
\end{itemize}

In this work, we show that the RM effect is negligible for atmospheric characterization at low spectral resolution. However, this conclusion is subtler at medium spectral resolution as there exist systems for which the RM bias is within the error bars of the JWST and HST instruments, even though the stellar lines are not fully resolved. In particular, we highlight that stellar lines infamous for contamination at high resolution should not be systematically disregarded at medium resolution. Finally, we note that the RM effect is dominant for K- and earlier-type stars because of their high rotational velocity, while other stellar contaminations may impact the transmission spectrum for slow rotators such as M stars (e.g., stellar spots; see \citealt{Genest2022}).
In the era of the JWST and upcoming ELT, our results highlight the need for low- and especially medium-resolution observations to be complemented by high-resolution measurements in order to disentangle atmospheric features from stellar contamination.

\begin{acknowledgements}
The authors warmly thanks L. Nortmann and E. Nagel for providing the near-infrared stellar template of GJ\,436. The authors also thank G. Morello for his insights on his methodology. This work has been carried out within the framework of the NCCR PlanetS supported by the Swiss National Science Foundation under grants 51NF40$\_$182901 and 51NF40$\_$205606. This project has received funding from the European Research Council (ERC) under the European Union's Horizon 2020 research and innovation programme (project {\sc Spice Dune}, grant agreement No 947634). This project has received funding from the European Research Council (ERC) under the European Union’s Horizon 2020 research and innovation program (grant agreement No 742095 ; SPIDI : Star-Planets-Inner Disk-Interactions; \url{http://www.spidi-eu.org}). This work has made use of the VALD database, operated at Uppsala University, the Institute of Astronomy RAS in Moscow, and the University of Vienna  

\end{acknowledgements}

%
%

\bibliographystyle{aa}
\bibliography{biblio}

\end{document}